\documentstyle[fake]{article}
\def\he{$^3$He}
\def\hep{$^3$He$^+$}

\def\crat{$^{12}{\rm C}/^{13}{\rm C}$}
\def\orat{$^{18}{\rm O}/^{16}{\rm O}$}
\def\mc{$M_{\rm CS}$}
\def\msun{$M_{\odot}$}

\def\lum{$\log L/L_{\odot}$}
\def\tef{$\log T_{\rm eff}$}
\begin{document}

\title{$^3$He in Planetary Nebulae:\\ a Challenge to Stellar Evolution Models}

\author{Daniele Galli\altaffilmark{1}, Letizia Stanghellini\altaffilmark{2}, 
Monica Tosi\altaffilmark{2} and Francesco Palla\altaffilmark{1}}

\altaffiltext{1}{Osservatorio Astrofisico di Arcetri, Largo E.~Fermi 5,
 I--50125 Firenze, Italy}

\altaffiltext{2}{Osservatorio Astronomico di Bologna,
 via Zamboni 33, I--40126 Bologna, Italy}

\authoremail{galli@arcetri.astro.it, stanghellini@astbo3.bo.astro.it, 
tosi@astbo3.bo.astro.it, palla@arcetri.astro.it}

\begin{abstract}

The discrepancy between the observed abundances of \he\ in the
interstellar medium and those predicted by stellar and galactic
chemical evolution remains largely unexplained.  In this paper, we
attempt to shed some light on this unsolved problem by presenting a
quantitative comparison of the \he\ abundances recently measured in six
planetary nebulae (IC~289, NGC~3242, NGC~6543, NGC~6720, NGC~7009,
NGC~7662) with the corresponding predictions of stellar evolution
theory.  The determination of the mass of the planetary
nebulae progenitors allows us to dismiss, to a good degree of
confidence, the hypothesis that the abundance of \he\ in the envelope
of all low--mass stars ($M\lesssim 2.5$~\msun) is strongly reduced with 
respect to the standard
theoretical values by some mixing mechanism acting in the latest phases
of stellar evolution.  The abundance versus mass correlation, allowance
made for the limitation of the sample, is in fact found to be fully
consistent with the classical prediction of stellar evolution.  We
examine the implications of this result on the galactic evolution of
\he\ with the help of a series of models with standard and
non--standard (i.e. \he\ depleted) nucleosynthesis prescriptions in
varying percentages of low--mass stars. The results are found to be
consistent with the abundances determined in the pre--solar material
and in the local interstellar medium {\it only} if the vast majority of
low--mass stars (more than 70--80~\%) follows non--standard
prescriptions. This implies that either the sample of planetary nebulae
under exam is highly biased and therefore not representative of the
whole population of low--mass stars, or the solution to the
\he\ problem lies elsewhere.

\end{abstract}


\section{Introduction}

Can the observed abundances of \he\ be used to set bounds on the
standard Big Bang nucleosynthesis models (SBBN)? The issue has been a
hot topic for many years, but the answer is still under debate. If the
\he\ abundance increases with time in the Galaxy, then the lowest
observed abundance of this isotope places a {\it lower} limit to the
baryon--to--photon ratio $\eta$ (see e.g. Yang et al.~1984).  Steigman
\& Tosi (1992), Vangioni-Flam, Olive \& Prantzos~(1994), and other
authors have shown that, neglecting the stellar production of \he, the
models of galactic chemical evolution of light isotopes can be safely
used to infer the primordial abundances, and hence to test the SBBN
predictions. On the other hand, starting from the classical papers by
Iben~(1967) and Truran \& Cameron~(1971), stellar models have always
predicted that low--mass stars are strong producers of \he. When
stellar production is included in models for the chemical evolution
of the Galaxy, no agreement between observed and predicted \he\ abundances
can be found (Rood, Steigman \& Tinsley 1976; and, more recently, Galli
et al.~1995, Olive et al.~1995, Dearborn, Steigman \& Tosi~1996,
hereafter DST, Fields~1996, Prantzos~1996).  Thus, the usefulness of
the helium isotope as a ``cosmological baryometer'' remains highly
questionable.

The strongest piece of evidence in favor of \he\ production in stars
comes from the observation of this isotope in galactic planetary
nebulae (PNs).  The first detection of \he\ in a PN (NCG~3242, by Rood,
Bania, \& Wilson~1992) confirmed the theoretical prediction that
low--mass stars are net producers of \he. The derived abundance (\he/H
$\sim 10^{-3}$) was found to be in good quantitative agreement with the
predicted values for stars of about one solar mass.  Subsequently, the
\he\ has been searched for in other PNs, and detected in two other
nebulae whose abundances are similar to that of NGC~3242 (see Rood et
al.~1995, hereafter RBWB, for an updated review).

The abundance of \he\ in PNs is much higher than that measured in the
solar system and in the interstellar medium (ISM) by one to two orders
of magnitude. This evidence is the heart of the so-called
``\he-problem''.  Indeed, observations toward a sample of galactic \ion{H}{2}
regions (Balser et al.~1994) and the very recent measurements by the
Ulysses spacecraft in the local interstellar cloud (Gloeckler \&
Geiss~1996) indicate a value of \he/H $\sim 10^{-5}$ .  It appears that
the ISM, today as well as at the time of the formation of the Sun, has
not been contaminated  by the high fractional abundances of
\he\ observed in the PNs studied by RBWB.  Since chemical abundances in
PNs are expected to represent the yields of low--mass stars, the
question is  whether the RBWB sample consists of an exception rather
than a rule in the evolution of low--mass stars.

In view of the importance of this argument, we have critically
analyzed  the problem in the sense of determining reliably the masses
of the PN progenitors in the RBWB sample, to check the observed
\he\ abundance to mass correlation against the stellar models.  The
basic knowledge of low--mass stellar evolution and of plasma
diagnostics allow us to tackle this basic problem in a quantitative
way.  

The outline of the paper is the following:  after a short review of the
most recent attempts to solve the \he\ problem at a stellar level
(Section~2), we determine the progenitor masses of the RBWB sample by
using up--to--date observed physical parameters of PNs (Section~3).  In
Section~4 we discuss the main steps for the calculations of the
\he\ abundances and we compare the resulting values with the
theoretical predictions of stellar models.  Finally, in Section~5 we
examine the statistical significance of the presently available
measurements in PNs in the context of detailed models of galactic
chemical evolution.

\section{Destruction of $^3$He in Low--Mass Stars}

In an influential paper, Hogan~(1995) established a relation between
the hypothetical mechanism responsible for the destruction of \he\ in
stellar envelopes, and the observed anomalies of the carbon isotopic
ratio (\crat) in a number of evolved stars. In fact, in some stars
(field giants, stars in galactic clusters) \crat\ falls below the
predictions of adequate standard models, but only so for masses below
$\sim 2$~\msun\ (Charbonnel~1994, Fig.~6).  These anomalously low
ratios can be accounted for by an extra-mixing process occurring after
the completion of the first dredge-up, and before the end of the Red
Giant Branch (RGB) phase.  It is reasonable to think that this process,
when present, would drastically alter the post-dredge-up envelope
abundance of other fragile isotopes, including \he.

Following this suggestion, Charbonnel~(1994, 1995) found that stellar
models of 0.8 and 1~\msun\ with extra-mixing nicely reproduce the low
\crat\ ratios in giant stars, the variation of lithium abundance
observed in Pop.~II evolved stars, and show a considerable destruction
of the envelope \he\ abundance.  However, this mechanism can operate
efficiently only in stars of mass up to $\sim 2$~\msun. The reason is
that the occurrence of the mixing process corresponds to the encounter
between the advancing hydrogen-burning shell and the discontinuity in
chemical composition left beyond by the convective envelope during the
dredge-up phase. As mixing cannot penetrate in a region of strong
molecular gradient, only after this evolutionary point, and only for
those stars where it can actually happen, trace elements in the
envelope (e.g. \he) can be transported down to the hydrogen burning
zone and, vice versa, freshly produced elements (e.g. $^{13}$C) can be
mixed in the convective region up to the stellar surface. For stars
more massive than  2~\msun\ the hydrogen-burning shell never reaches the
chemically homogeneous region.

Wasserburg, Boothroyd \& Sackmann~(1995) provided additional support to
the extra-mixing hypothesis by computing a set of stellar models for
RGB and Asymptotic Giant Branch (AGB) stars undergoing Cool Bottom
Processing (CBP), in which an ad hoc mixing mechanism transports
stellar fluid elements from the cool bottom of the convective envelope
down to some inner layer hot enough for nuclear processing, and
vice versa.  The model results reproduce the anomalous \crat\ observed
in low--mass RGB stars and, at the same time, show a large destruction
of envelope \he\ (by a factor $\sim 10$ in a 1~\msun\ star).  The same
process can also resolve the puzzling low \orat\ ratios observed in AGB
stars (see also Sackmann \& Boothroyd~1996 and Boothroyd \&
Malaney~1996).  Similar results have been obtained by Denissenkov \&
Weiss~(1996) and Weiss, Wagenhuber \& Denissenkov~(1996), who model
deep mixing as a diffusion process and show that a number of observed
surface abondance correlations can be quantitatively reproduced,
although with slightly different sets of mixing parameters in each
case.

In conclusion, non-standard mixing on RGB and/or AGB phases offers an
attractive scenario whose basic features can be summarized as follows:
({\it i}\/) contrary to the standard view, stars of mass less than
2~\msun\ destroy \he\ during their post-MS evolution and return \he\
depleted material to the ISM, consistently with the abundance of \he\
measured in the pre-solar material, in galactic \ion{H}{2} regions, and in the
local ISM; ({\it ii}\/) the \crat\ ratios measured in the envelopes of
giants less massive than 2~\msun, the \orat\ ratios in the envelopes of
AGB stars of the same mass range, and the lithium abundances in metal-poor
giants can
be quantitatively reproduced; ({\it iii}\/) high abundances of \he\ are
allowed in those PNs whose progenitor mass is larger than 2~\msun\ or
did not otherwise undergo extra--mixing.  Thus, the \he\ puzzle is
solved at a {\it stellar} level.

The question to address now is then: do the PNs with known
\he\ abundance comply with the scenario outlined above?  The answer
can only come from an analysis of the mass of their progenitor stars.

\section{Planetary Nebulae: the Masses of Progenitor Stars}

Central stars (CSs) of PNs are the relics of the AGB stars that have
gone through the instability driven ejection of the envelope, after the
thermal pulse phase.  The progenitor mass is then the stellar remnant
mass plus the mass ejected during post--Main Sequence (post--MS)
evolution. The direct measure of the ejected mass is not possible, and
the study of the CS is not always very straightforward, given that the
star is hidden by the surrounding nebula. We determine the progenitor
masses through the following observational and interpretative steps:

({\it i}\/) We firstly derive the CS's  \ion{He}{2} Zanstra temperature, with
the method described in Kaler~(1983). When assuming that this value is
a good approximation of the effective temperature, we make two basic
assumptions:  the strongest one is that we assimilate the stellar
output to a black--body spectrum, the weakest one is to assume that the
nebula is thick to the  He I--ionizing radiation.  In order to
calculate Zanstra temperatures we need the values of the H$\beta$ and
the \ion{He}{2} ($\lambda$4686) fluxes, corrected for the atmospheric
extinction. Also needed are the angular diameters.  We take these basic
parameters from the catalog of Cahn, Kaler \& Stanghellini (1992,
hereafter CKS), which represents the most complete and recent
compilation of these PN data.  We then use the $V$ and $B$ magnitudes
as quoted in Acker et al. (1992, hereafter A92), and, where possible,
we use averages of the \ion{He}{2} Zanstra temperature derived from the $B$
and $V$ magnitudes, to minimize the errors on the data sets.

({\it ii}\/) To proceed in our analysis, we need an estimate of the
distances to the PNs.  The first approach is to use statistical distance
from CKS. These distances have been derived with the assumption that
the ionized mass of PNs is the same for all optically thin PNs, and is
a function of the surface brightness for optically thick PNs. These are
obviously strong assumptions and result in large distance errors,
although the CKS distances are considered among the most
reliable statistical distances in the literature, given the very
careful calibration (see Terzian~1993). If other (non--statistical)
distances are available, we take them into account.  The stellar
luminosity is then evaluated from the \ion{He}{2} temperature and the most
reliable distance available.

({\it iii}\/) The derivation of the stellar mass from the observed
luminosity and temperature is obtained by placing the star on the $\log
L$--\tef\ diagram (see Figure~1), and read off the mass by comparing its
position with the synthetic evolutionary tracks  (Stanghellini \&
Renzini~1993). It is worth noting that these tracks have been
calculated for H-burning CS, while there are indications that some PN
nuclei are H-depleted, thus their energy is supplied by He--burning
(see a recent review in Stanghellini 1995).  For He--burning stars, the
tracks would have similar shapes on the $\log L$--\tef\ diagram to
those of H--burning stars, and the mass derivation should not be very
different.  The only possible caveat concerns stars that burn hydrogen
at the AGB, then the H-burning ceases and they switch to
helium--burning.  Only in this very special case the track on the $\log
L$--\tef\ diagram experiences a blue loop, thus the luminosity at a
given mass, and for each value of the effective temperature, is not
exactly the same than that of H--burning stars. Since the evolutionary
timescales of these blue loops are quite short in comparison with the
overall post--AGB evolution, we conclude that our mass determination
based on the usual H--burning tracks is reliable.

({\it iv}\/) In order to calculate MS masses from the CS masses
discussed above, we  use the empirical initial mass--final mass
relation derived by Weidemann (1987) from observation of field white
dwarfs. Given that this  method is not model--dependent, we prefer
this  approach instead of using models of post--AGB evolution for our
calculations.

In the following sections we illustrate the post--AGB and progenitor
mass derivation of the RBWB sample PNs.  In Table 1 we list the usual
PN name, the size in arcsec, the statistical and individual distances
in kpc, the \ion{He}{2} Zanstra temperature, the luminosity, the CS
mass, the progenitor mass calculated with the statistical (index [s]),
and the individual distance (index [i]). The best mass and distance
determinations are in boldface. Table~1 clearly shows that the PNs of
the sample have masses lower than $\sim 2$~\msun, NGC~6720 being a
marginal case.  In such a range, the proposed extra-mixing mechanism,
active prior to the PN ejection, should have effectively destroyed all
the \he.

\subsection{IC~289 }

IC~289 is an irregular multiple shell PN (Chu, Jacoby \& Arendt~1987)
whose CS magnitudes are known in terms of lower limits only ($B>15.1$,
$V>15.9$, Shaw \& Kaler 1985). By using these lower limits, we produce
upper limits to the \ion{He}{2} Zanstra temperature and luminosity.  The
statistical distance is $D({\rm s})=1.43$~kpc (CKS), and the individual
distance is $D({\rm i})=2.71$~kpc (Kaler \& Lutz~1985).  We obtain \tef
$=4.965 \pm 0.034$, and \lum $=3.51 \pm 0.16$ by using the
statistical distance, and \lum $=4.06 \pm 0.16$ with the individual
distance. The errors associated with the Zanstra analysis depend on the
intrinsic observing uncertainties of the individual measurements.
By placing the CS of IC~289 on the $\log L$--\tef\ diagram we obtain
\mc(s) $< 0.58$~\msun\ or \mc(i) $< 0.75$~\msun\ depending on the
distance scale used.  We should use the statistical distance as a prime
indicator, since the wind distances can be overestimated (Kaler 1991,
priv. comm.).  The empirical initial mass-final mass relation yields
$M_{\rm MS}({\rm s})=1.64$~\msun, if the CS mass is calculated with the
statistical distance.

\subsection{NGC~3242 }

The physical parameters of this multiple shell, attached halo PN have
been extensively discussed in Stanghellini \& Pasquali~(1995).  We thus
will not repeat here the analysis that has been performed to obtain the
post--AGB mass (\mc=0.56~\msun\ with $D({\rm s})=0.88$~kpc). The MS
mass, calculated through the empirical initial mass--final mass
relation, is $M_{\rm MS}=(1.2 \pm $0.2)~\msun. The individual distances
available for this nebula, quoted in A92, show a large spread ($<D({\rm
i})>=0.9\pm 1$~kpc), and we do not use them in our calculations. Recently,
an expansion distance measured with radio observations (Hajian,
Phillips \& Terzian~1995) places this PN at $0.42\pm 0.16$~kpc. The
stellar luminosity at this distance will drop down to $\log
L/L_\odot=2.671$, and the corresponding post-AGB mass would lie out of
the permitted range.

\subsection{NGC~6543 }

This H-rich WR nucleus (Mend\'ez~1991) has well-determined magnitudes
(A92). The \ion{He}{2} and H$\beta$ fluxes are from A92, while the
angular diameter is quoted in CKS. We obtain \tef$=4.854 \pm 0.015$ and
\lum $=3.547 \pm 0.081$ with CKS statistical distance. The only
non-statistical distance available to NGC~6543 is the wind distance by
Kaler \& Lutz~(1985), which is very close to the the statistical
distance.  Other values of \tef\ found in the literature are within
10~\% of our value (see e.g.  Bianchi, Recillas \& Grewing~1989,
Perinotto~1993, Castor et al.~1981).  The position on the HR diagram
yields a CS mass of \mc $=(0.58 \pm 0.01)$~\msun, which, compared to
the initial mass--final mass empirical relation (Weidemann~1987) gives
$M_{\rm MS}({\rm s})=(1.6 \pm 0.2)$~\msun.

\subsection{NGC~6720 }

NGC~6720 also hosts a H-rich nucleus (Mend\'ez~1991).  Its statistical
distance is $D({\rm s})=0.87$~kpc (CKS), and the measured expansion
distance is 0.5~kpc (Pottasch~1980).  By using the fluxes and angular
dimension of CKS and the magnitudes quoted in A92, we find \tef $=5.148
\pm 0.026$ and \lum=2.858 $\pm 0.070$, which translates into a mass of
\mc $=(0.61 \pm 0.03)$~\msun.  If we use the expansion distance we
obtain \lum$ =2.375 \pm 0.070$, which pushes the stellar mass up to
0.69~\msun. This second value is also in agreement with the derivation
of $M_{\rm V}=7.3$ determined by Pier et al. (1993), although this last
result is quite fragile. The statistical distance is in agreement with
that found by Napiwotzki \& Sch\"onberner~(1995), $D=0.99$~kpc. Very
recently, Manchado et al.~(1996) have found that this nebula has a
second, attached shell. If we were to use this second diameter to
calculate the statistical distance {\it \`a la} CKS, we would have obtained
$D=0.5$~kpc. In conclusion, it is difficult to decide which is the best
guess for the distance.  We calculate the mass of the progenitor to be
$M_{\rm MS}({\rm s})=(2.2\pm 0.6)$~\msun\ and $M_{\rm MS}({\rm
i})=3.8$~\msun, respectively, but it is clear that more accurate
measurements (e.g. with radio expansion velocity) are needed for this
object.

\subsection{NGC~7009 } \vskip5pt

Another hydrogen--rich nucleus (O(H), Mend\'ez~1991). The fact that
several PNs of our sample have H--rich nuclei is important in that
their mass determination from the H-burning post--AGB tracks are very
reliable. Its distance measure is controversial:  while the statistical
distance places it at 1.2~kpc (CKS), the individual distance values
quoted in A92 (except the wind and the model-dependent measurements)
average to about half of this value. We calculate \tef\ via Zanstra
analysis by using the fluxes and dimensions of CKS, and find
\tef$=4.965 \pm 0.017$; by using the statistical distance we obtain
\lum $=3.41 \pm 0.10$.  The effective temperature is well in
agreement with other values found in the literature (Pottasch~1993,
Heap~1993, Perinotto~1993). If we were to calculate the luminosity with
$D({\rm i})=0.5$~kpc we would have found \lum $=2.646$.  The first \lum
value gives a mass of $M_{\rm CS}({\rm s})=(0.57 \pm 0.01)$~\msun,
while the other calculated luminosity is too low to allow a mass
determination.  Other authors find slightly higher values for the mass
(e.g. \mc $\simeq 0.64$, Heap~1993). The larger distance is supported
also by a very recent work by Maciel~(1995), who quotes $D=1.6$~kpc
derived via UV and radio kinematics. For the progenitor mass we obtain
$M_{\rm MS}({\rm s})=(1.4\pm 0.2)$~\msun; a progenitor mass cannot be
given for $D=0.5$~kpc.

\subsection{NGC~7662 }

The magnitude determinations for this CS are rather poor (A92). We
calculate the effective temperature and luminosity using the fluxes and
dimensions of CKS.  We find \tef $=5.067 \pm 0.071$ and \lum $=3.44
\pm 0.18$ by using the statistical distance (CKS), and \lum $=3.06 \pm
0.18$ with the averaged individual distance (A92).  As a result,
\mc(s) $=(0.59 \pm 0.02)$~\msun, or \mc(i) $=0.56$~\msun\ with
similar uncertainties. With our mass determination we find $M_{\rm
MS}({\rm s})=(1.7 \pm 0.3)$~\msun\ and $M_{\rm MS}({\rm i})=1.2$~\msun,
respectively. Recently, Hajian \& Terzian~(1996) find a radio expansion
distance of $D=0.79\pm 0.75$~kpc, in good agreement with the value used
here ($D=0.75$~kpc) and within the range of the statistical distance
given by CKS. In conclusion, we calculate the progenitor mass using
$D=0.75$~kpc.

\section{The Abundance of $^3$He in Planetary Nebulae}

In order to derive the \he\ abundance we model PNs as homogeneous
spheres of fully ionized gas.  We compute the abundance of \he\ in PNs
from the line parameters given by RBWB and from our analysis of the PNs
physical parameters.  

\subsection{The Density of $^3${\rm He}$^+$}

The \hep\  column density in PNs can be obtained from
observations of the hyperfine structure line of \hep\ at
$\nu=8.6656$~GHz:
$$
N(^3{\rm He}^+)={g_l+g_u\over g_u}{8\pi k\nu^2\over hc^3A_{ul}}
\int T_B(v) {\rm d}v,
\eqno(1)
$$
where $g_u=1$, $g_l=3$, $A_{ul}=1.95436\times 10^{-12}$~s$^{-1}$ (Gould~1994),
and $T_B(v)$ is the brightness temperature profile of the line.
For a gaussian line profile,
$$
\int T_B {\rm d}v={1\over 2}\sqrt{{\pi\over \ln 2}}T_B^0 \Delta v,
\eqno(2)
$$
where $T_B^0$ is the brightness temperature at the center of the line and
$\Delta v$ is the full width at half power.

The brightness temperature $T_B^0$ is related to the observed beam-averaged
brightness temperature $T_L$, given by RBWB, by:
$$
T_B^0=T_L{\theta_b^2+\theta_s^2\over \theta_s^2},
\eqno(3)
$$
where $\theta_b$ and $\theta_s$ are the beam and source angular radii,
respectively ($2\theta_b=82\arcsec$).

The number of \hep\ atoms per unit volume $n(^3{\rm He}^+)$ can be
obtained dividing the column density $N(^3{\rm He}^+)$ by the average
optical path $<\Delta s>$ through the source. Representing a PN as a
homogeneous sphere of radius $R=\theta_s D$, the optical path at a
position angle $\theta$ is $\Delta s(\theta)=2\sqrt{R^2-(\theta
D)^2}$, and the optical path averaged on the source results $<\Delta
s>=\pi R/2 =\pi\theta_s D/2$.

The final expression for $n(^3{\rm He}^+)$ is then
$$
n(^3{\rm He}^+)=8\sqrt{{\pi\over \ln 2}}{g_l+g_u\over g_u}
{k\nu^2\over hc^3A_{ul}} 
{\theta_b^2+\theta_s^2\over D\theta_s^3} T_L \Delta v.
\eqno(4)
$$
Inserting numerical values, we obtain
$$
n(^3{\rm He}^+)=22.7 
\left({T_L\over {\rm mK}}\right)
\left({\Delta v\over {\rm km}~{\rm s}^{-1}}\right)
\left({D\over {\rm kpc}}\right)^{-1}
\left({\theta_s\over \arcsec}\right)^{-3}
\left({\theta_b\over 41\arcsec}\right)^2
\left(1+{\theta_s^2\over \theta_b^2}\right)~{\rm cm}^{-3}.
\eqno(5)
$$

\subsection{The Density of {\rm H}$^+$}

For a ionized gas containing H$^+$, He$^+$ and He$^{2+}$, the density of H$^+$ is
related to the density of electrons via 
$$
n({\rm H}^+)={n({\rm e})\over 1+y(1+x)},
$$
where 
$$
y={n({\rm He}^+)+n({\rm He}^{2+})\over n({\rm H}^+)},~~~x={n({\rm He}^{2+})\over 
n({\rm He}^+)+n({\rm He}^{2+})}.
$$
The values of $y$, $x$ (from CKS) and $n({\rm e})$ are shown in columns
2 to 4 of Table~2. With the exception of IC~289, the electronic
densities are derived from forbidden line intensities (Stanghellini \&
Kaler~1989).  The values shown are the averages of the mean values for
each density indicator.  No forbidden line data being available for
IC~289, we have computed its electronic density from the radio flux at
5~GHz with the help of the formula given by Gathier~(1987):
$$ n({\rm e})=4.96\times 10^3
\left({S_{5\,{\rm GHz}}\over {\rm mJy}}\right)^{1/2}
\left({T_{\rm e}\over 10^4\;{\rm K}}\right)^{1/4}
\left({D\over {\rm kpc}}\right)^{-1}
\left({\theta_s\over \arcsec}\right)^{-3/2}\epsilon^{1/2}
\left[{1+y(1+x)\over {1+y(1+3x)}}\right]^{1/2}\;{\rm cm}^{-3},
\eqno(6)
$$
where we have taken $S_{5\,{\rm GHz}}=212$~mJy (Higgs~1971), the
electron temperature $T_{\rm e}=1.55\times 10^4$~K from CKS, the
average $\theta_s$ from Table~1, and a filling factor $\epsilon=1$.  We
estimate an uncertainty of 10~\% on the resulting values of $n({\rm
H}^+)$.

The \he\ line parameters from RBWB and our derived abundances are
listed in the last four columns of Table~2. Given the distance and the
line parameters, the resulting range of abundances reflects mainly the
uncertainty in the angular radius of the source (see eq.~[5]), which we
have allowed to vary between the minimun and maximum inner radius of
the nebula as given by Chu et al.~(1987).  Our results agree with those
of RBWB.  The only exception is for IC~289, for which RBWB obtain a
value of $11.6\times 10^{-4}$, outside our range. We cannot identify
the source of the discrepancy since the values of the physical
parameters adopted for each PN by RBWB are not given. In agreement with
RBWB we conclude that the present analysis confirms the fact that the
ejecta of stars with masses below $\sim 2.5$~\msun\ have abundances a
factor 10--100 larger than those observed in the solar system and in
the local ISM.

\subsection{Comparison with Stellar Evolutionary Models}

Having determined the \he\ abundance in the six PNs of known progenitor
mass, we can now compare these values with the predictions of stellar
evolution models. In Figure~2 we show the derived values of the
\he\ abundance as function of the stellar mass. Going from the values
listed in Table~2 to those plotted here, we have assumed that \he/H$
\simeq$ \hep/H$^+$ (Balser et al.~1994). The boxes represent the
uncertainty associated with the mass and \he\ abundance
determinations.  For IC~289 we have also assumed a lower limit of
$0.8$~\msun\ for the progenitor mass.  For the PNs with upper limits on
the abundance, the uncertainty in the progenitor mass is indicated by
the size of the horizontal bar.  The predictions of several stellar
evolution models have been considered. From Figure~2 we see that the
most recent calculations agree very well with each other, whereas
Iben's results give higher \he\ abundance at each mass. The source of
this discrepancy lies on  Iben's underestimate of the \he\ destruction
cross section (see Galli et al.~1995). In any case, all PNs with
measured abundance are fully consistent with the theoretical
predictions.

In Figure~2 we also show the expected \he\ abundance in the case of
non-standard mixing. The most extensive calculations are those of
Boothroyd~(1996), while Hogan~(1995) only gives a crude
estimate of the equilibrium abundance independent of mass. The detailed
calculations indicate that the mass dependence of the destruction of
\he\ is quite strong, and the resulting abundance decreases sharply for
masses below $\sim 2.5$~\msun. The comparison with the observed PNs
clearly shows that these stars have not suffered any depletion.  Even
for the most massive progenitor (NGC~6720), for which the difference
between the two cases is smaller, the observed abundance is still a
factor $\sim 2$ above the non-standard curve. We thus conclude that the
current observations do not support the conjecture of enhanced
\he\ depletion in all low--mass stars.

\section{Chemical Evolution of $^3$He}

As discussed in Section 1, chemical evolution models adopting standard
$^3$He stellar nucleosynthesis overproduce $^3$He.  In particular, all
the galactic models in better agreement with the observational
constraints predict $^3$He abundances largely inconsistent with those
observed in the solar system and in the ISM (locally and at different
galactic radii), unless they adopt alternative nucleosyntheses with
strongly reduced $^3$He contribution from low and intermediate mass
stars of the kind described in Section 2 (see Tosi~1996 and references
therein). On the other hand, if all stars with $M\leq 2.5$~\msun
were to deplete their envelope $^3$He down to a mass fraction
$X_3\simeq 1\times 10^{-5}$, no PN would be able to show abundances 100
times larger as those observed by RBWB. By the same argument, the
possibility of a nuclear physics solution to the \he\ problem proposed
by Galli et al.~(1994) should be dismissed.

A possible way out of this inconsistency is that some stars experience
the extra-mixing and deplete $^3$He and some others do not and maintain
the high yield predicted by standard nucleosynthesis models.  In order
to verify whether or not this suggestion can reconcile the galactic
requirements with the high $^3$He abundances of RBWB's PNs, we have
computed a series of chemical evolution models with standard and
alternative nucleosynthesis prescriptions in varying percentages of low
and intermediate mass stars.

To this aim, we have recomputed some of the numerical models discussed
by DST.  All the results described here refer to DST's model 1, a model
consistent with all the major observational constraints of the disk
(Tosi 1988a,b, Giovagnoli \& Tosi~1995). This model assumes an
exponentially decreasing SFR (with e-folding time 15 Gyr), explicitly
dependent on both the gas and total mass density currently observed at
each galactocentric distance, a constant (in time), uniform (in space)
infall rate of 0.004 M$_{\odot}$~kpc$^{-2}$~yr$^{-1}$ and Tinsley's
(1980) initial mass function.  For consistency with DST findings on the
deuterium evolution, the metallicity of the infalling gas is not
primordial and assumed to be 1/5 of solar.  The sun is assumed to be
located at 8 kpc from the galactic center and to have formed 4.5 Gyr
ago. The current disk age is assumed to be 13 Gyr (but see DST for the
modest effect of assuming instead an age of 10 Gyr).  Based on Tosi's
(1996) comparison of the best chemical evolution models currently
available in the literature, here we adopt $X_{2,p} = 5\times 10^{-5}$
and $X_{3,p} = 2\times 10^{-5}$ as the primordial abundances by mass of
deuterium and $^3$He, respectively.

For the cases with standard stellar nucleosynthesis we have adopted
DST's yields; for the cases with $^3$He depletion induced by
extra-mixing we have alternatively adopted either Boothroyd's (1996)
detailed values as function of stellar mass, or simply taken the
equilibrium value \he/H=1$\times 10^{-5}$ for $M< 2.5$~\msun\ as in
Hogan's (1995) suggestion.

Let us define $P_d$ as the percentage of stars with $M\leq 2.5$~\msun\
experiencing extra--mixing and therefore depleting $^3$He; the
remaining $1-P_d$ fraction of stars have standard $^3$He yields.
Figure~3 shows the evolution in the solar ring of the \he/H ratio
resulting from assuming $P_d=0$ (DST standard model
1-C-Ib), $P_d =$ 0.7, 0.8, 0.9 and Boothroyd's (1996) yields. The
vertical bars correspond to the 2-$\sigma$ ranges of values derived
from observations of the solar system and the local ISM (Geiss 1993 and
Gloeckler \& Geiss 1996, respectively).  It is apparent that only with
$P_d\leq$ 0.7 can the models fit the observed ranges.  The long-dashed
curve in the Figure shows the effect of assuming $P_d = 0.8$ and
Hogan's depletion.  Since the latter is more drastic than Boothroyd's,
the percentage of depleting stars required to obtain the same agreement
with the data is smaller, but the results are qualitatively the same.

Figure~4 shows the $^3$He/H radial distributions resulting at the
present epoch from the models shown in Figure~3. Also shown are the
abundances derived by RBWB from \ion{H}{2} region radio observations (dots
with their error bars) and by Gloeckler \& Geiss~(1996) from Ulysses
data on local ISM (vertical bar for the 2-$\sigma$ range). To get a
radial distribution flat and low enough to fit the data, $P_d$ values
larger than 0.7 must be invoked, in agreement with the results obtained
for the local evolution.

We thus suggest that to solve the \he\ problem in terms of extra mixing
in low and intermediate mass stars, the vast majority of them must be
affected by this phenomenon. In this framework, the few PNs observed by
RBWB and showing large \he\ content must have been selected in the,
relatively small, sample of stars without deep mixing. Indeed, the
selection criteria for the target PNs (Rood 1996, priv.  comm.) were
aimed at maximizing the likelihood of detecting the \he\ line:  ({\it
i}\/) located at least 500 pc above the galactic plane; ({\it ii}\/)
medium excitation PNs; ({\it iii}\/) PNs with low nitrogen and $^{13}$C
to avoid objects where mixing could have destroyed \he. The latter
criterion suggests that the possibility that 70-80\% of low--mass stars
deplete \he\ and the remaning 30-20\% do not is a viable solution to
the \he\ problem.

%An alternative scenario has been proposed by Olive et al.~(1995) who
%dismissed the possibility that extra mixing could reduce the \he\ yield
%of low--mass stars on the basis of the observational data for PNs, and
%favored a {\it galactic} solution to the \he\ problem, in which winds
%from massive stars (largely depleted in \he) pollute and dilute the
%ionized gas of \ion{H}{2} regions, in a larger extent for the more massive
%\ion{H}{2} regions. The pre-solar \he\ abundance might be accounted for by
%the same process, if the solar system formed in an OB association.
%However, according to Scully et al.~(1996) in this case the dilution of
%\he\ in the pre-solar material would only be of the order of $\sim
%10$\%, otherwise supernovae exploded prior to the formation of the
%solar system would largely overproduce O and Ne. Such a tiny dilution
%cannot solve the \he\ problem.  Clearly, this {\it galactic} solution
%to the \he\ problem needs to be worked out in detail and quantitatively
%tested against observations before being accepted or dismissed.  It
%offers the appealing characteristic of being fully consistent with the
%\he\ data in PNs and with standard models of stellar evolution, but can
%hardly account for Gloeckler \& Geiss' (1996) low \he\ in the local
%ISM.

\section{Conclusions}

We have analyzed the sample of PNs with measured \he\ abundance in
order to determine the mass of the progenitor stars. We found that all
PNs have masses below $\sim 2.5$~\msun, and their observed abundances
are in agreement with the predictions of standard nucleosynthesis in
low--mass stars. Unless the PN sample of RBWB is confirmed to be
highly biased in favour of non-depleting stars, these results would
pose severe problems to the non-standard destruction mechanisms
recently suggested in order to overcome the long-standing problem of
\he\ overproduction on the Galactic timescale.

By using models of galactic evolution of \he\ with standard and
non-standard nucleosynthesis prescriptions, we have found that the
resulting evolution of \he\ can be consistent with the values
determined in the pre-solar material and in the local ISM only if more
than 70--80~\% of the whole population of stars with mass below $\sim
2.5$~\msun\ has undergone enhanced \he\ depletion. This implies that
either the sample of PNs studied here is not representative of the
low--mass stellar population, or the solution to the \he\ problem lies
elsewhere. As for the former possibility, a crucial observational test
would be the simultaneous determination of the \he\ abundance and the
\crat\ ratio in a large sample of PNs.  In fact, in addition to the
\he\ depletion, extra-mixing during the AGB phase would also decrease
the \crat\ ratio from to the standard value of $\simeq 30$ to $\simeq
5$ (see e.g. Sackmann \& Boothroyd~1996). Such small values have been
observed in few PNs (Bachiller et al.~1996), although for the only PN
(NGC~6720) of known \he\ abundance the \crat\ is consistent with
standard predictions.  Extending this kind of observations to a
statistically significant number of PNs will shed new light on the long
standing problem of \he.

Given the size and the selection criteria mentioned in the previous
Section, the RBWB sample of PNs is too small and selective to draw firm
conclusions about the generality of the depletion processes taking
place in the latest stages of stellar evolution. Observations of the
same kind, but on a much larger sample, including also {\it depleting
candidates}, i.e.  PNs with high nitrogen and \crat\ are necessary to
finally understand both the late evolutionary phases of low--mass stars
and the galactic evolution of an important cosmological baryometer like
\he.

\acknowledgements 

It is a pleasure to thank Dr. A. Boothroyd for providing the
extra--mixing yields prior to publication; Dr. C. Charbonnel for her
careful reading of the manuscript and valuable comments; Dr. R. T. Rood
for useful discussions on the \he\ measurements in PNs. D.G. wishes to
thank the Institute for Nuclear Theory at the University of Washington
for its hospitality and the Department of Energy for partial support
during the completion of this work.  L.S. acknowledges the warm
hospitality of the STScI where part of this work was carried out.

\clearpage 

\begin{deluxetable}{lllllllllll}

\scriptsize
\tablewidth{0pt}
\tablecaption{Distances, Temperatures, Luminosities and Masses of Central
Stars, and Progenitor Masses}
\tablehead{
\colhead{ PN name } & 
\colhead{$ \stackrel{\textstyle \theta }{ (\arcsec) }^{({\rm a})} $} & 
\colhead{$ \stackrel{\textstyle D({\rm s}) }{ ({\rm kpc}) } $} & 
\colhead{$ \stackrel{\textstyle D({\rm i}) }{ ({\rm kpc}) } $} & 
\colhead{$ \log \frac{T_{\rm eff}}{\rm K}^{({\rm b})} $} & 
\colhead{$ \log \frac{L({\rm s})}{L_\odot} $} & 
\colhead{$ \log \frac{L({\rm i})}{L_\odot} $} & 
\colhead{$ \stackrel{\textstyle M_{\rm CS}({\rm s}) }{ (M_\odot) } $} & 
\colhead{$ \stackrel{\textstyle M_{\rm CS}({\rm i}) }{ (M_\odot) } $} & 
\colhead{$ \stackrel{\textstyle M_{\rm MS}({\rm s}) }{ (M_\odot) } $} & 
\colhead{$ \stackrel{\textstyle M_{\rm MS}({\rm i}) }{ (M_\odot) } $}  
}
\startdata

IC~289 & 18 & {\bf 1.43} & 2.71 & $<4.965\pm 0.034$ & $<3.51\pm 0.16$ 
& $<4.060$ & $<0.58$ & $<0.75$ & $<$ {\bf 1.6} & $<4.7$ \nl
 
NGC~3242 & 16 & {\bf 0.88} & 0.42 & $4.963\pm 0.009$ & $3.318\pm 0.030$ 
& 2.671 & $0.56\pm 0.01$ & --$^{\rm (d)}$ & {\bf 1.2 $\pm$ 0.2} & --$^{\rm (d)}$ \nl
 
NGC~6543 & 9.4 & {\bf 0.98} & 0.89 & $4.854\pm 0.015$ & $3.547\pm 0.081$ 
& 3.461 & $0.58\pm 0.01$ & 0.57 & {\bf 1.6 $\pm$ 0.2} & 1.4 \nl
 
NGC~6720 & 34 & {\bf 0.87} & 0.5 & $5.148\pm 0.026$ & $2.858\pm 0.070$ 
& 2.375 & $0.61\pm 0.03$ & 0.69 & {\bf 2.2 $\pm$ 0.6} & 3.8 \nl
 
NGC~7009 & 14 & {\bf 1.2} & 0.5$^{\rm (e)}$ & $4.965\pm 0.017$ & 
$3.41\pm 0.10$ & 2.646 &
$0.57\pm 0.01$ & --$^{\rm (d)}$ & {\bf 1.4 $\pm$ 0.2} & --$^{\rm (d)}$ \nl
 
NGC~7662 & 7 & 1.16 & {\bf 0.75}$^{\rm (e)}$ & $5.067\pm 0.071$ & $3.44\pm 0.18$ 
& 3.060 & $0.59\pm 0.02$ & 0.56 & $1.7\pm 0.3$ & {\bf 1.2} \nl
  
\enddata
\tablenotetext{(a)}{from CKS}
\tablenotetext{(b)}{He {\sc II} Zanstra temperatures}
\tablenotetext{(c)}{same uncertainty as in $\log L({\rm s})$}
\tablenotetext{(d)}{off permitted range}
\tablenotetext{(e)}{average values: $\sigma_{\rm NGC7009}=0.08$; $\sigma_{\rm NGC7662}=0.32$}

\end{deluxetable}

\clearpage

\begin{deluxetable}{lllllllll}

\footnotesize
\tablewidth{0pt}
\tablecaption{Abundances in PNs}
\tablehead{
\colhead{PN name}  &
\colhead{$y^{\rm (a)}$} &
\colhead{$x^{\rm (a)}$} &
\colhead{$ \stackrel{\textstyle n({\rm e})^{\rm (b)} }{ ({\rm cm}^{-3}) } $} & 
\colhead{$ \stackrel{\textstyle n({\rm H}^+)     }{ ({\rm cm}^{-3})   } $} & 
\colhead{$ \stackrel{\textstyle T_L^{\rm (c)}    }{ ({\rm mK})        } $} & 
\colhead{$ \stackrel{\textstyle \Delta v^{\rm (c)} }{ ({\rm km s}^{-1})} $} & 
\colhead{$ \stackrel{\textstyle n(^3{\rm He}^+)    }{ ({\rm cm}^{-3})    } $} & 
\colhead{$ \stackrel{\textstyle ^3{\rm He}^+/{\rm H}^+ }{ \times 10^4  } $} 
}

\startdata

IC~289   & 0.110 & 0.448  & $7.34\times 10^2$ & $6.33\times 10^2$ 
         & 2.75  &  37.73  &  0.25--0.55  & {\bf 3.9--8.7} \nl
 
NGC~3242 & 0.100 & 0.187 & $2.69\times 10^3$ & $2.40\times 10^3$
         & 4.16  &  47.24  &  1.6--5.4  &  {\bf 6.7--22} \nl
 
NGC~6543 & 0.111 & 0.000 & $1.95\times 10^3$ & $1.76\times 10^3$
         & $<$4.82  & 56.50  & $<$(5.8--9.1) &  {\bf $<$(33--52)} 
	 \nl
 
NGC~6720 & 0.113 & 0.231 & $4.90\times 10^2$ & $4.30\times 10^2$
         & 2.85  & 36.55 & 0.066-0.13 & {\bf 1.5--3.0} \nl
 
NGC~7009 & 0.112 & 0.113 & $4.07\times 10^3$ & $3.62\times 10^3$
         & $<$3.64 & 43.82 & $<$(2.1--11.3) & {\bf $<$(5.8--31)}  
	 \nl
 
NGC~7662 & 0.107 & 0.287 & $2.63\times 10^3$ & $2.31\times 10^3$
         & $<$6.98 & 40.73 & $<$(9.1--26) & {\bf $<$(39--112)}  
	 \nl

\enddata

\tablenotetext{(a)}{from CKS}
\tablenotetext{(b)}{from Stanghellini \& Kaler~(1989), except IC~289 (see text)}
\tablenotetext{(c)}{from RBWB}

\end{deluxetable}

\clearpage

\figcaption{Location of the six PNs of the RBWB sample in the H-R
diagram. Solid lines are evolutionary tracks from Stanghellini \&
Renzini~(1993) for central stars with masses 0.55, 0.57, 0.58, 0.59,
and 0.60~\msun (bottom to top). \label{fig1}}

\figcaption{\he\ abundance (by number with respect to H) versus MS mass
for the six PNs of the RBWB sample with the associated range of derived
values.  The curves show the results of the \he\ abundance in the
stellar envelope at the end of the RGB phase as computed by:
Iben~(1967) for $Z=0.02$ ({\it solid line}); Rood et al.~(1976) for
$Z=0.02$ ({\it short dashed line}); Boothroyd~(1996) for $Z=0.02$ and
$X_{3,{\rm MS}}=8.4\times 10^{-5}$ ({\it long dashed line}); Weiss et
al.~(1996) for $Z=0.02$ and $X_{3,{\rm MS}}=6.02\times 10^{-5}$ ({\it
dot dashed line}); DST for $Z=0.02$ and $X_{3,{\rm MS}}=1.0\times
10^{-4}$ ({\it dotted line}).  The results of stellar nucleosynthesis
ith deep mixing during the RGB phase computed by Boothroyd~(1996) are
shown as {\it dot-dashed} lines, and the equilibrium value \he/H
$=10^{-5}$ for $M<2.5$~\msun\ as a {\it short dash - long dash} line.
\label{fig2}}

\figcaption{Time evolution of \he/H in the solar neighborhood.  The
vertical bars (2-$\sigma$ errors) show the abundance of \he\ measured in
the solar system (Geiss 1993) and in the local ISM (Geiss \& Gloeckler
1996).  The curves show the predictions of chemical evolution models
assuming different percentages $P_d$ of stars depleting \he\ (see
text):  {\it solid line} $P_d$=0 with DST standard stellar yields ({\it
solid line}); $P_d$=0.7, 0.8 and 0.9 with Boothroyd's~(1996) yields
({\it dash-dotted, dotted {\rm and} dashed lines}); $P_d=0.8$ with
Hogan's~(1995) depleted yields ({\it long-dashed line}). \label{fig3}}
 
\figcaption{Radial distribution of \he/H as derived from H~II region
observations (dots and error bars from RBWB) and from chemical
evolution models for the present epoch.  The vertical bar gives (at
2-$\sigma$) the value measured in the local ISM (Geiss \& Gloeckler
1996). The line symbols are as in Figure~3. \label{fig4}}
 
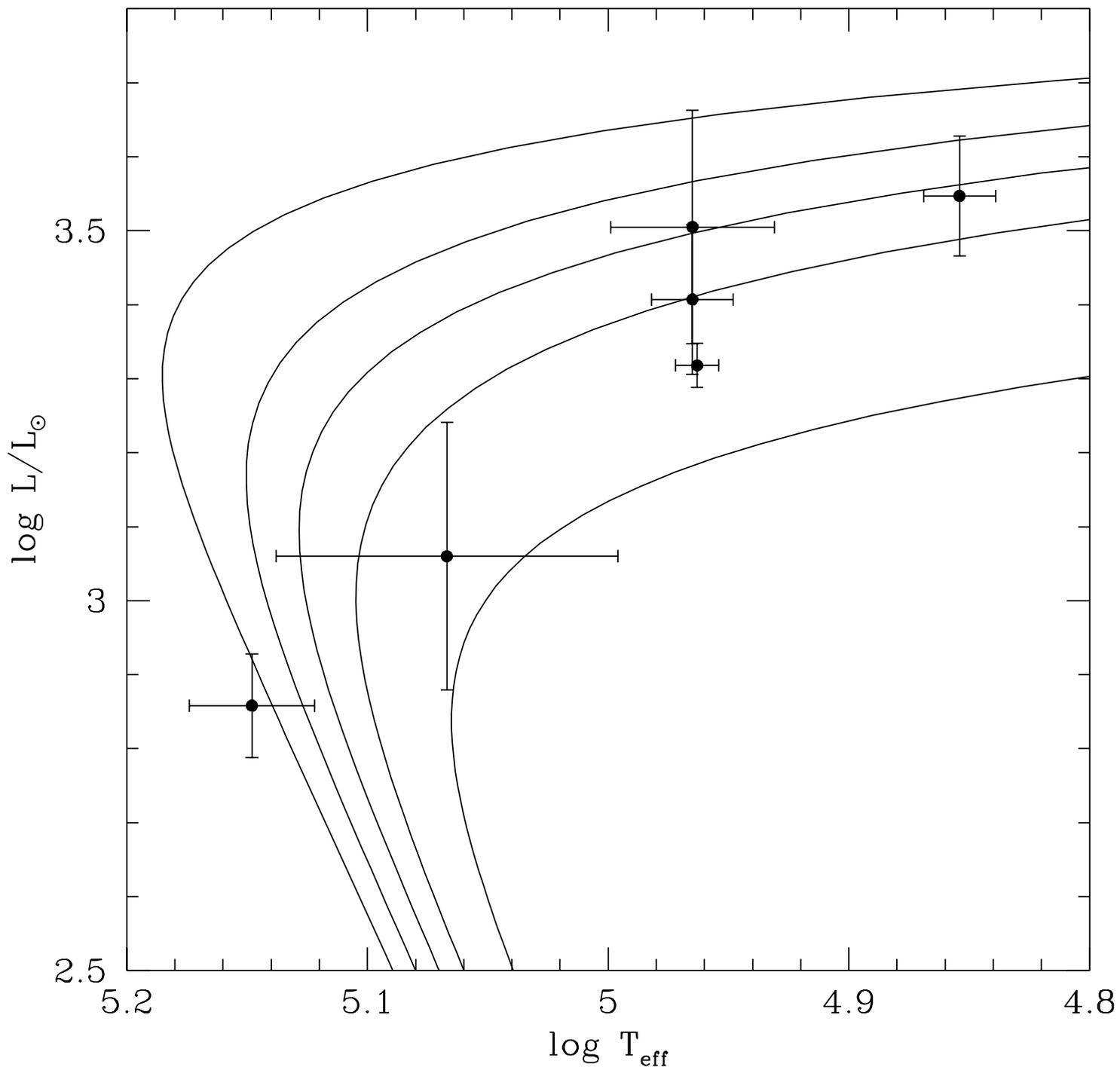
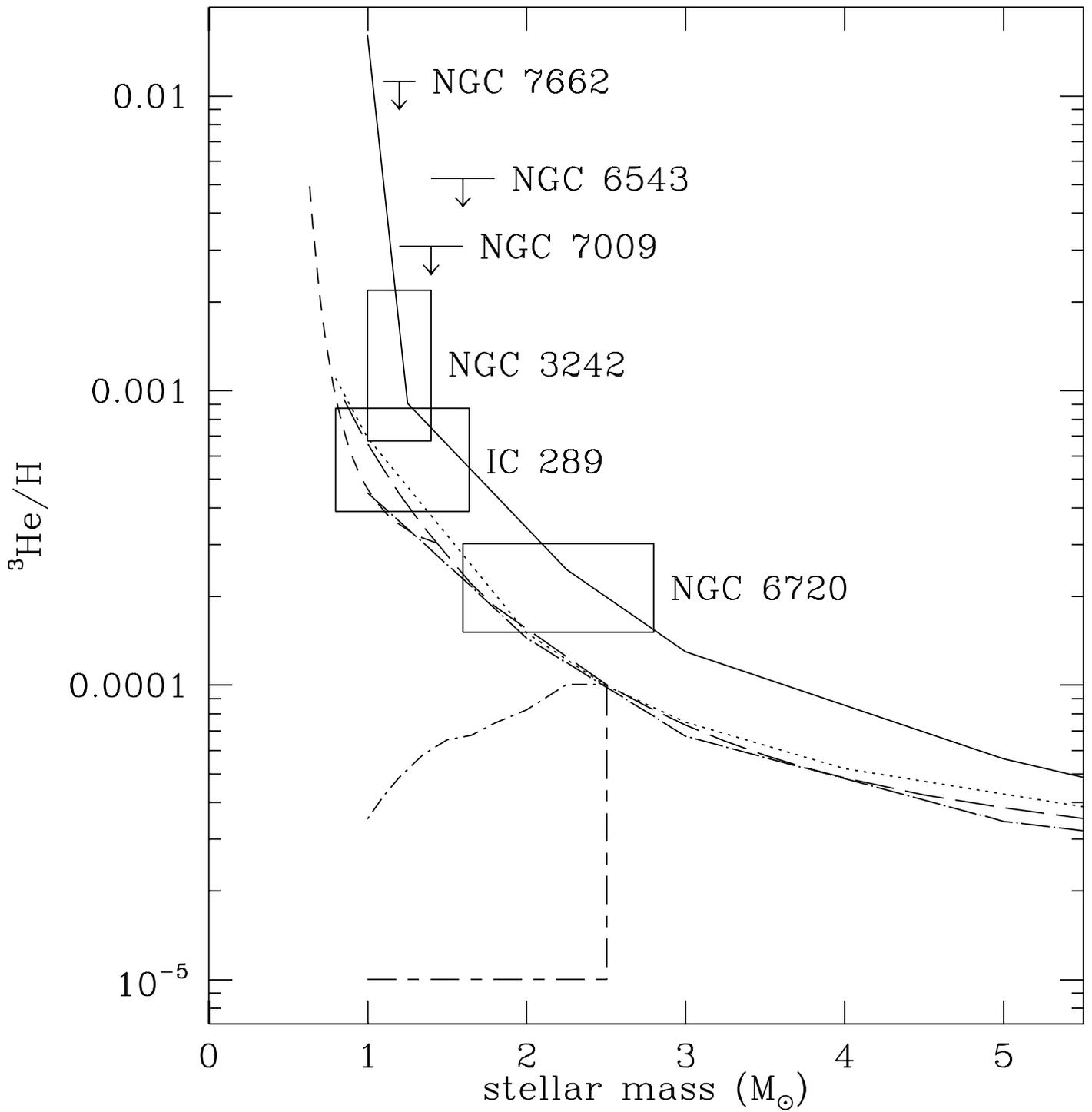
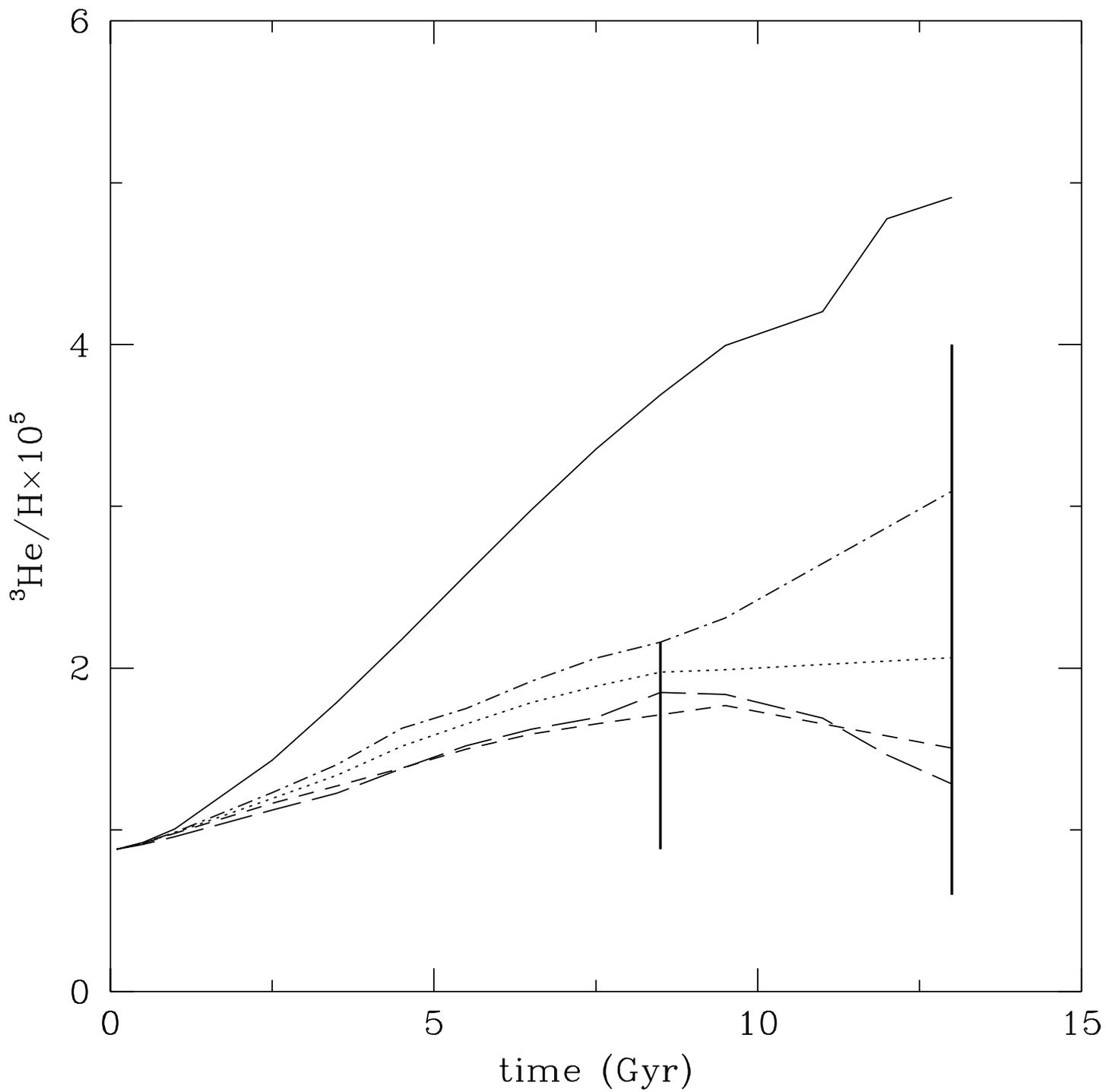
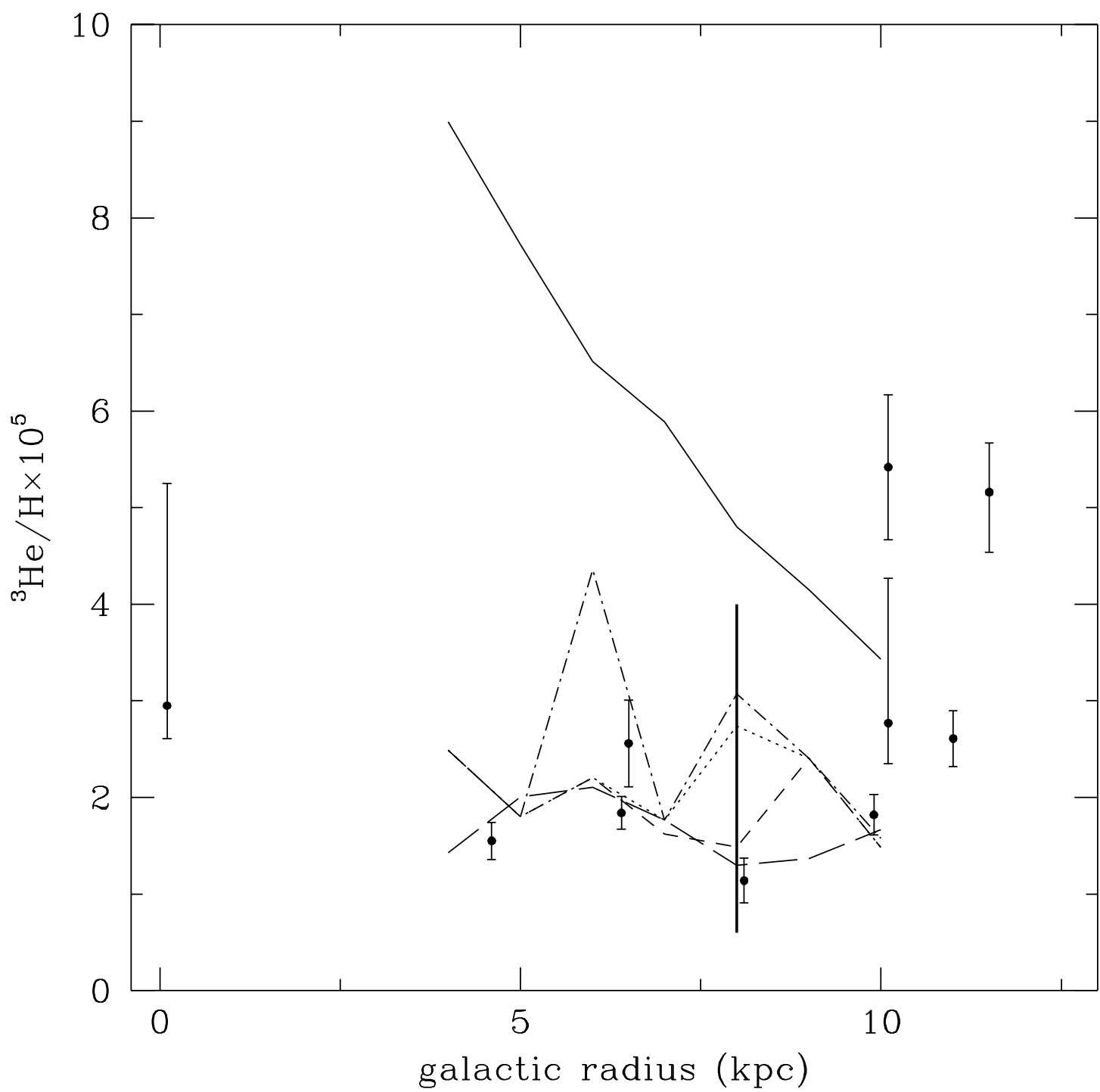
\end{document}